\documentclass[twocolumn,showpacs,preprintnumbers,prb,fleqn]{revtex4}
\usepackage{epsfig}
\usepackage{graphicx}
\usepackage{amsfonts}
\begin{document}
                                                                                

\title{Hole and electron dynamics in the triangular-lattice antiferromagnet --- \\
interplay of frustration and spin fluctuations}
\author{Pooja Srivastava}
\email{psriv@iitk.ac.in}
\author{Avinash Singh}
\email{avinas@iitk.ac.in}
\affiliation{Department of Physics, Indian Institute of Technology Kanpur - 208016}
\date{\today}
\begin{abstract}
Single-particle dynamics in the 120$^{\circ}$ ordered antiferromagnetic state 
of the triangular-lattice Hubbard model is studied using a physically transparent
fluctuation approach in terms of multiple magnon emission and absorption processes
within the noncrossing approximation.
Hole and electron spectral features are evaluated at intermediate $U$,
and analyzed in terms of a competition between the frustration-induced direct hopping 
and the virtual hopping. 
Finite $U$-induced competing interactions and frustration effects contributing through 
the magnon dispersion are also discussed. 
Finite contribution to self-energy correction from long-wavelength (Goldstone) modes, 
together with the high density of electron scattering states in the narrow, sharp peak 
in the upper band, result in strong fermion-magnon scattering 
leading to pronounced incoherent behaviour in the electron dynamics.
The fluctuation-induced first-order metal-insulator transition 
due to vanishing band gap is also discussed. 
\end{abstract}
\maketitle

\section{Introduction}
There has been renewed interest in correlated electron systems on triangular lattices,
as evidenced by recent studies of antiferromagnetism, superconductivity 
and metal-insulator transition in the organic systems 
$\rm \kappa -(BEDT-TTF)_2 X$,\cite{review1,review2} 
the discovery of superconductivity in $\rm Na_x Co O_2 . y H_2 O$,\cite{watersup}
the observation of low-temperature insulating phases 
in some $\sqrt{3}$-adlayer structures such as K on Si[111],\cite{weitering} 
and quasi two-dimensional $120^\circ$ spin ordering and spin-wave excitations
in $\rm Rb Fe (MoO_4)_2$ (Refs. 5,6) 
and the multiferroic material $\rm Ho Mn O_3$.\cite{holmium,sw}

The recent finding of finite $U$-induced competing interactions and frustration 
in the $120^\circ$ ordered antiferromagnetic (AF) state of the Hubbard model on a triangular lattice,\cite{tri} resulting in vanishing spin stiffness at $U \approx 6$
and a magnetic instability towards a F-AF state at $U \approx 7$, 
both in the insulating state, adds a new dimension to the intrinsic 
geometric frustration of the triangular-lattice antiferromagnet.
Indeed, strongly enhanced quantum spin fluctuations associated with the magnetic instability may account for 
the absence of long-range magnetic order in the {\em nearly isotropic} 
organic antiferromagnet $\rm \kappa -(BEDT-TTF)_2 Cu_2  (CN)_3$,
as inferred from recent $^1$H NMR and static susceptibility measurements down to 32 mK, 
well below the estimated exchange constant $J \sim 250$ K,
suggesting the realization of a quantum spin-liquid state.\cite{kanoda}
The realization of a non-magnetic insulator at intermediate $U$ is also interesting 
as it allows, with decreasing $U$, for a Mott-type metal-insulator transition 
not accompanied by any magnetic symmetry breaking, 
as seen in the layered system $\rm \kappa -(BEDT-TTF)_2 Cu[N(CN)_2]Cl$,\cite{kagawa}
and currently of much theoretical interest.\cite{pirg2,pirg3}

Recently, 
self-energy corrections due to multiple magnon processes
in the AF state of the frustrated square-lattice $t-t'$-Hubbard model
were evaluated using a fluctuation approach.\cite{self} 
Quasiparticle dispersion obtained with the same set of Hubbard model cuprate parameters
as obtained from a recent magnon spectrum fit\cite{spectrum}
was found to yield excellent agreement with ARPES data for 
$\rm Sr_2 Cu O_2 Cl_2$,\cite{self} thus providing a unified description 
of magnetic and electronic excitations in cuprates.

It is therefore interesting to examine and contrast self-energy corrections
and quasiparticle behaviour in the $120^\circ$ ordered AF state of the triangular-lattice Hubbard model which involves non-collinear ordering, intrinsic geometric frustration,
and also finite $U$-induced competing interactions and frustration. 
Indeed, we find that long-wavelength magnon modes yield 
finite contribution to the fermion-magnon scattering process, 
unlike the square-lattice case where this contribution was 
negligible.\cite{self} 

Frustration and spin fluctuations are involved in an interesting interplay
with respect to quasiparticle behaviour.
As neighbouring spins are not antiparallel in
triangular-lattice and frustrated square-lattice AF states,
frustration results in an O($t$) or O($t'$) direct hopping in addition 
to the O($J$) virtual hopping. 
Competition between the two dispersions results in band broadening/narrowing, 
which has a dramatic effect on self-energy corrections 
due to significantly different density of fermion scattering states. 
Competition also results in a reduced band gap,
thus bringing the system closer to a metal-insulator (MI) transition. 
The $t-J$ model calculations do not involve this competition 
as the virtual hopping term is absent.

In this paper, we examine the hole and electron spectral functions in terms of
self-energy corrections in the $120^\circ$ ordered AF state of the Hubbard model
on a triangular lattice.
Hole and electron dynamics in an antiferromagnetic background is associated with 
multiple magnon emission and absorption processes corresponding to 
broken AF bonds, string states, and scrambling of the AF spin ordering. 
The fluctuation approach adopted in this paper in terms of a 
diagrammatical expansion provides a physically transparent picture 
in which the hole motion is renormalized due to the fluctuating transverse 
field associated with spin fluctuations.
This fluctuation perspective is physically relevant 
in view of the substantial ordered-moment reduction,
which approaches $59\%$ in the strong-coupling (Heisenberg) limit.\cite{capriolti}

Single-particle dynamics in the AF state of the triangular-lattice $t-J$ model 
has been studied in the self-consistent Born approximation 
(SCBA),\cite{dombre,trumper,apel} exact diagonalization,\cite{dombre}    
and projection techniques.\cite{vojta,dagotto}
One-electron density of states
has been examined using the quantum Monte Carlo method,\cite{qmc_green}
showing a pseudogap development for intermediate $U$.

We consider the Hubbard model 
\begin{equation}
{\cal H} =  -t \sum_{i\delta\sigma} a_{i,\sigma} ^\dagger a_{i+\delta,\sigma} 
+ U \sum_i a_{i\uparrow} ^\dagger a_{i\uparrow} a_{i\downarrow} ^\dagger a_{i\downarrow}
\end{equation}
with nearest-neighbour (NN) hopping on a triangular lattice.
The model has particle-hole symmetry under the transformation $t \rightarrow -t$.
In this paper we consider the case of positive $t$ and hole (electron) 
doping in the lower (upper) band;
same result holds for negative $t$ and electron (hole) doping in the upper (lower) band.
In the following we set the hopping energy $t=1$. 

The organization of this paper is as follows.
The three-sublattice basis is briefly reviewed in section II 
to introduce the notation and key features of 
the classical-level fermion dispersion.
Transverse spin fluctuations are introduced in section III 
in terms of magnon amplitudes and energies. 
Intraband self-energy corrections due to multi-magnon processes
are studied in section IV in the non-crossing approximation.
Hole and electron spectral functions for the lower and upper bands 
are discussed in section V and conclusions are presented in section VI.

\section{Three-sublattice representation}
While the spiral-state description applies only to Bravais lattices,
the sublattice-basis description applies to Kagom\'{e} type non-Bravais 
lattices as well. In general, the $120^\circ$ AF state is characterized by an 
ordering plane (normal $\hat{n}_1$) and a planar direction ($\hat{n}_2$) in 
spin space, with reference to which spin orientations are given by
\begin{equation}
\hat{\alpha} = \cos\phi_\alpha \; \hat{n}_2
+ \sin\phi_\alpha (\hat{n}_1 \times \hat{n}_2 )
\end{equation}
corresponding to angles $\phi_\alpha=0^\circ$, $120^\circ$, and $240^\circ$
on the three sublattices $\alpha=\rm A,B,C$.
A convenient choice is $\hat{n}_1=\hat{z}$ 
(spin-ordering in the $x-y$ plane) and $\hat{n}_2=\hat{x}$, 
so that the local mean field ${\bf \Delta}_\alpha = \frac{1}{2}
U\langle \Psi_i ^\dagger {\mbox{\boldmath $\sigma$}} \Psi_i \rangle_\alpha $
in the $120^\circ$ ordered Hartree-Fock state is given by
\begin{equation}
{\bf \Delta}_\alpha = \Delta \hat{\alpha} \;\;\;\;\; (\hat{\alpha}=\hat{a},\hat{b},\hat{c})
\end{equation}
on the three sublattices
in terms of the three lattice unit vectors 
\begin{equation}
\hat{a} = \hat{x}, \;\;\;\; 
\hat{b} = -\frac{1}{2}\hat{x} + \frac{\sqrt{3}}{2} \hat{y}, 
\;\;\;\;\;
\hat{c} = -\frac{1}{2}\hat{x} - \frac{\sqrt{3}}{2} \hat{y} .
\end{equation}

\begin{figure}
\vspace*{-22mm}
\hspace*{-20mm}
\psfig{figure=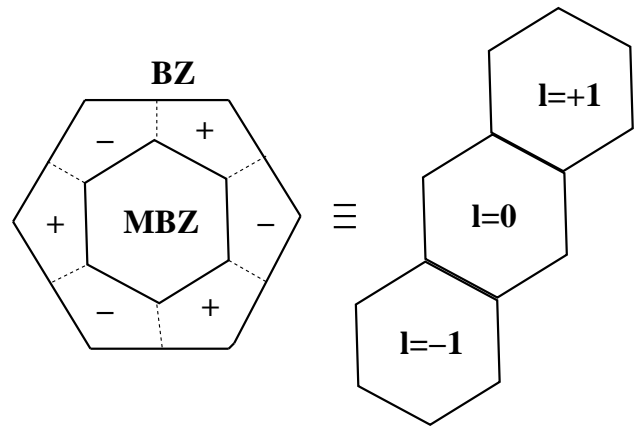,width=100mm,angle=-90}
\vspace{-20mm}
\caption{Brillouin zone (BZ) for the triangular lattice 
and magnetic Brillouin zone (MBZ) for the $120^\circ$ ordered AF state,
along with the BZ - MBZ correspondence involving the branch index $l$.}
\end{figure}

Fourier transformation within the sublattice basis yields
\begin{equation}
{\cal H}_{\rm HF} = \sum_{\bf k} \Psi_{\bf k} ^\dagger
\left [ \begin{array}{ccc}
-{\mbox{\boldmath $\sigma$}}.{\bf \Delta}_{\rm A} & \delta_{\bf k} & 
\delta_{\bf k}^*  \\
 & & \\
\delta_{\bf k}^* & - {\mbox{\boldmath $\sigma$}}.{\bf \Delta}_{\rm B} & 
\delta_{\bf k} \\
 & & \\
\delta_{\bf k}  & \delta_{\bf k}^* & 
-{\mbox{\boldmath $\sigma$}}.{\bf \Delta}_{\rm C} 
\end{array}
\right ]\Psi_{\bf k} \; ,
\end{equation}
where $\Psi_{\bf k} \equiv 
(a_{{\bf k}\uparrow}\;a_{{\bf k}\downarrow}
\; b_{{\bf k}\uparrow}\;b_{{\bf k}\downarrow}
\; c_{{\bf k}\uparrow}\;c_{{\bf k}\downarrow} )$ in terms of the fermion operators 
$a_{\bf k},b_{\bf k},c_{\bf k}$ defined on the three sublattices A, B, C. 
Wavevector $\bf k$ lies within the Magnetic Brillouin Zone (MBZ),
corresponding to the three inter-penetrating triangular sublattices (lattice parameter $\sqrt{3}a$). 
The NN hopping term 
\begin{equation}
\delta_{\bf k} = -t \sum_{\hat{\delta}=\hat{a},\hat{b},\hat{c}}
 e^{i{\bf k}.\hat{\delta} }
= -t [e^{i k_x} + 2e^{-ik_x/2} \cos (\sqrt{3} k_y/2) ]
\end{equation} 
mixes AB, BC, and CA sublattices, which are connected by the three lattice
unit vectors. The lattice hopping term $\delta_{\bf k}$ yields the 
triangular-lattice free-fermion energy
\begin{equation}
\epsilon_{\bf k} = \delta_{\bf k} + \delta_{\bf k} ^\ast 
\end{equation}
and transforms as 
\begin{equation}
\delta_{\bf k\pm Q} = 
\delta_{\bf k} \, e^{\pm i 2\pi/3} 
\end{equation} 
under momentum translation by the ordering wavevector 
${\bf Q}=(2\pi/3,2\pi/\sqrt{3})$. 

The $[6\times 6]$ Hamiltonian matrix obeys the cyclic property 
$[{\cal H}_{\rm HF}]_{AB} = [{\cal H}_{\rm HF}]_{BC} =
[{\cal H}_{\rm HF}]_{CA}$ of the $120^\circ$ ordered state,
resulting in the following spin-sublattice structure  of the normalized eigenvectors
\begin{equation}
|{\bf k},l \rangle = \frac{1}{\sqrt{3}} \left ( \begin{array}{l} 
\alpha_{{\bf k},l} \; e^{-i\phi_\alpha} \\
\beta_{{\bf k},l} \;  \end{array} \right )_\sigma
\otimes \left ( 
\begin{array}{c} 1 \\ e^{i\theta_l} \\ e^{-i\theta_l} \end{array} \right )_\alpha \; ,
\end{equation} 
where the planar spin orientations $\phi_\alpha = 0^\circ,\; 120^\circ,\; 240^\circ$ 
for the three sublattices, 
and the sublattice phase angle $\theta_l = 2\pi l/3 $ for the three 
fermion branches $l=0,\pm 1$. 
Substituting the above eigenvector structure, 
and contracting over the sublattice sector $\alpha$,
the eigenvalue equations of the $[6\times 6]$ Hamiltonian matrix 
reduce to three $[2\times 2]$ equations in the spin sector corresponding to $l=0,\pm 1$;
the six eigenvalues $E_{{\bf k},l}^\mp$ and amplitudes 
$(\alpha_{{\bf k},l}^\mp \; \beta_{{\bf k},l}^\mp )$ 
for the lower ($-$) and upper ($+$) AF bands are then given by
\begin{eqnarray}
E_{{\bf k},l}^\mp &=& \eta_{\bf k} \pm \sqrt{\Delta^2 + \xi_{\bf k}^2} \\
\alpha_{{\bf k},l}^\mp &=& \pm \frac{1}{\sqrt{2}}\left (1 \pm 
\frac{\xi_{\bf k}}{\sqrt{\Delta^2 + \xi_{\bf k}^2}} \right )^{1/2} \nonumber \\
\beta_{{\bf k},l}^\mp &=& + \frac{1}{\sqrt{2}}\left (1 \mp \frac{\xi_{\bf k}}{\sqrt{\Delta^2 + \xi_{\bf k}^2}} \right )^{1/2} \; ,
\end{eqnarray}
where
$\eta_{\bf k} \equiv (\epsilon_{\bf k} + \epsilon_{\bf k-Q})/2 $ and
$\xi_{\bf k} \equiv (\epsilon_{\bf k} - \epsilon_{\bf k-Q})/2$,
with momentum values $\bf k$, $\bf k \pm Q$ 
corresponding to fermion branches $l=0,\pm 1$, respectively.
The mean field $\Delta$ and magnetization $m$ are related to $U$ 
through the self-consistency condition.\cite{tri} 

\begin{figure}
\vspace*{-68mm}
\hspace*{-35mm}
\psfig{figure=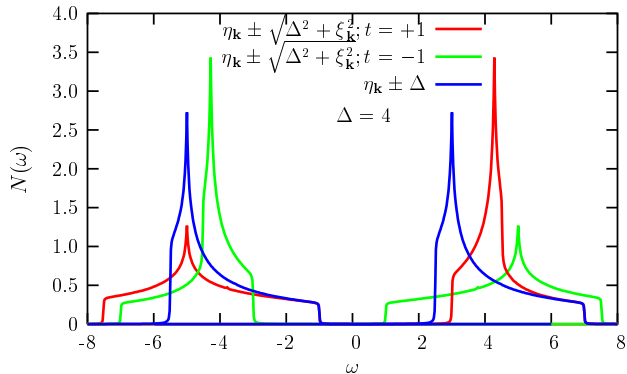,width=140mm}
\vspace{-82mm}
\caption{(color online) Comparison of density of states corresponding to the 
full HF dispersion and the direct hopping dispersion,
showing the band broadening (narrowing) due to the virtual hopping dispersion.}
\end{figure}
                                                                                
Typical of the AF state,
upper band states yield a negative contribution to spin densities 
due to the negative sign of $\alpha_{\bf k} ^+$.
Consequently, when the two bands start overlapping on decreasing $U$, 
partial occupation of the upper band has an amplified effect on reduction of 
magnetic order --- 
reduced sublattice magnetization due to band overlap decreases the mean-field $\Delta$, which further increases the overlap. Typically, the magnetic order 
therefore falls very rapidly after band overlap sets in. 

The density of states (DOS) corresponding to the full HF (classical) dispersion in Eq. (10),
shown in Fig. 2 for both positive and negative $t$,
exhibits a key competition between the two dispersion terms associated with 
direct hopping $(\eta_{\bf k})$ of order $t$ 
and virtual hopping $(\sqrt{\Delta^2 + \xi^2 _{\bf k}} - \Delta)$,
which is of order $J$ in the strong-coupling limit. 
For positive $t$, both dispersion terms favour same state at the top of the lower band,
while they favour different states at the bottom of the upper band,
thereby causing band narrowing. 
The competition results in broadening and narrowing of the 
lower and upper bands, depending on the sign of $t$, 
which has a dramatic effect on the self-energy correction 
due to significantly different density of fermion scattering states. 
There is no such competition in the $t-J$ model studies where 
the virtual hopping dispersion term is absent. 

From Eq. (8), it follows that 
\begin{equation}
\epsilon_{\bf k} = -(\epsilon_{\bf k+Q} + \epsilon_{\bf k-Q})
\end{equation}
so that  
\begin{equation}
\eta_{\bf k-Q} = -\frac{\epsilon_{\bf k}}{2} \; .
\end{equation}
Therefore, the direct hopping dispersion  $\eta_{\bf k-Q}$,  
to which the HF quasiparticle dispersion (10) 
reduces in the strong-coupling limit,
is identical to the classical-level dispersion $\epsilon_{\bf k}/2$ taken in 
earlier $t-J$ model studies, corresponding to effective hopping 
$t\cos (120^\circ) = -t/2$ associated with the $120^\circ$ ordering,
where momentum translation by ${\bf Q}$ connects 
the real and slave fermions.\cite{dombre}
Comparison of DOS corresponding to the full dispersion with that of $\eta_{\bf k}$
is also shown in Fig. 2. 

\section{Transverse spin fluctuations}
Including both transverse and longitudinal spin fluctuations,
the full spin-fluctuation propagator $\langle \Psi_{\rm G} |{\rm T}[
S_i ^\mu (t) S_j ^\nu (t')]\Psi_{\rm G} \rangle$ 
in the $120^\circ$ ordered AF state
has been studied recently in the random phase approximation (RPA) 
in the full $U$ range.\cite{tri}
Even in the intermediate-coupling regime,
the magnitude of longitudinal fluctuation $\langle S_\alpha ^2 \rangle$ 
along the local ordering directions $\hat{\alpha}$ was found to be quite negligible
($\langle S_{\alpha} ^2 \rangle \sim 10^{-4}$ at $U \approx 7$). 
We therefore focus on transverse 
spin fluctuations along the two locally normal spin directions.

Spin rotation about the $\hat{z}$ direction
\begin{equation}
\left ( \begin{array}{c} \sigma_x \\ \sigma_y \\ \sigma_z \end{array} \right )'
= \left [ \begin {array}{ccc} \cos \phi_\alpha & \sin \phi_\alpha & 0 \\
-\sin \phi_\alpha & \cos\phi_\alpha & 0 \\
0 & 0 & 1 \end{array} \right ]
\left ( \begin{array}{c} \sigma_x  \\ \sigma_y \\ \sigma_z \end{array} \right )
\end{equation}
by angles
$\phi_\alpha=0^\circ,120^\circ,-120^\circ$ for $\alpha$ = A,B,C
renders $x'$ as the spin ordering direction for all three sublattices.
In the $2\otimes 3$ spin-sublattice basis of the two transverse spin directions
$\mu,\nu=y',z'$ and the three sublattices $\alpha,\beta=A,B,C$,
the RPA-level spin-fluctuation propagator is then given by
\begin{equation}
[\chi ({\bf q},\omega)]_{\alpha\beta} ^{\mu\nu}  = \frac{\frac{1}{2}\;[\chi^0 ({\bf q},\omega)]}
{{\bf 1} - U[\chi^0 ({\bf q},\omega)]} \; , 
\end{equation}
where the bare particle-hole propagator 
\begin{eqnarray}
[\chi^0({\bf q},\omega)]^{\mu\nu} _{\alpha \beta} =&& \frac{1}{2}
\sum_{{\bf k},l,m}   \left [ 
\frac{\langle \sigma_\mu \rangle_\alpha ^{-+}
\langle \sigma_\nu \rangle_\beta^{-+*}}
{E_{{\bf k-q},m}^+ - E_{{\bf k},l}^- + \omega} 
\right. \nonumber \\ 
 && \left. \; \; \;\; \; \;\; \; \;\; \; \; \;  
+\frac{\langle \sigma_\mu \rangle_\alpha ^{+-}
\langle \sigma_\nu \rangle_\beta ^{+-*}}
{E_{{\bf k},m}^+ - E_{{\bf k-q},l}^- - \omega} \right ]
\end{eqnarray}
involves integrating out the fermions in the broken-symmetry state.
In the particle-hole matrix elements of the rotated spins
\begin{equation}
\langle \sigma_\mu \rangle_\alpha ^{-+} \equiv
\langle {\bf k-q},m| \sigma_\mu |{\bf k},l\rangle_\alpha 
\end{equation}
the spin orientation angles $\phi_\alpha$ in the fermion states (Eq. 9)
are transformed out.

We now discuss the spin-sublattice structure of the $[\chi^0({\bf q},\omega)]$ matrix
and its eigenvectors.
While the spin-diagonal blocks $[\chi^0({\bf q},\omega)]^{\mu\mu}$ are Hermitian 
and the off-diagonal blocks $[\chi^0({\bf q},\omega)]^{\mu \bar{\mu}}$  
are anti-Hermitian, they obey the cyclic symmetry 
$[\chi^0({\bf q},\omega)]_{AB} = [\chi^0({\bf q},\omega)]_{BC} =
[\chi^0({\bf q},\omega)]_{CA}$ of the $120^\circ$ ordered phase. 
Also, the sublattice diagonal elements $[\chi^0({\bf q},\omega)]_{\alpha\alpha}$ 
are all identical due to sublattice symmetry. 
Consequently, the normalized eigenvectors of $[\chi^0({\bf q},\omega)]$
have the following spin-sublattice structure
\begin{equation}
|\phi _\lambda \rangle = \frac{1}{\sqrt{3}}
\left ( \begin{array}{r} -iu \\ v \end{array} \right )_\mu
\otimes \left ( 
\begin{array}{c} 1 \\ e^{i\,2\pi \lambda /3 } \\ e^{-i\,2\pi \lambda /3} \end{array} 
\right )_\alpha 
\end{equation} 
where $\lambda =0,\pm 1$  for the three magnon branches,
and the real and normalized amplitudes $u$ and $v$ represent the fluctuation 
amplitudes in the $y'$ and $z'$ directions, respectively.
Contracting over the sublattice index,
the eigenvalue equation for $| \phi \rangle$ therefore reduces to 
\begin{equation}
[\chi^0 ({\bf q},\omega)] | \phi \rangle_\lambda = 
[\chi^0 _\lambda({\bf q},\omega)] \left ( \begin{array}{c} u \\ v \end{array} \right )
= \lambda_{\bf q} \left ( \begin{array}{c} u \\ v \end{array} \right ) \; ,
\end{equation}
where $[\chi^0 _\lambda ({\bf q},\omega)]$ is a $[2\times 2]$ real-symmetric matrix. 

Solving the pole equation $1-U\lambda_{\bf q}(\omega)=0$ 
for the magnon energy $\omega_{\bf q}$,
and expanding $\lambda_{\bf q}(\omega)$ around the poles to  obtain the magnon amplitudes, 
yields the magnon propagator
\begin{equation}
[\chi ({\bf q},\omega)]  = \sum_{\lambda=0,\pm 1} \left [
\frac{| {\bf q},\lambda \rangle \langle {\bf q},\lambda |_A }
{\omega + \omega_{{\bf q},\lambda} - i\eta}
- 
\frac{| {\bf q},\lambda \rangle \langle {\bf q},\lambda |_R }
{\omega - \omega_{{\bf q},\lambda} + i\eta} \right ]  \; ,
\end{equation}
where the magnon eigenvectors for the advanced (A) and retarded (R) modes are given by
\begin{equation}
|{\bf q},\lambda \rangle = \frac{1}{\sqrt{3}}
\left ( \begin{array}{r} \mp i {\cal Y}_{{\bf q},\lambda} \\ 
{\cal Z}_{{\bf q},\lambda} \end{array} \right )_\mu
\otimes \left ( 
\begin{array}{c} 1 \\ e^{i\,2\pi \lambda /3 } \\ e^{-i\,2\pi \lambda /3} \end{array} 
\right )_\alpha 
\end{equation} 
in terms of the magnon amplitudes 
\begin{eqnarray}
&&{\cal Y}_{{\bf q},\lambda} = u_{{\bf q},\lambda} 
/\sqrt{2U^2 |d\lambda_{{\bf q},\lambda}/d\omega| } \;\;\;\;\;\;\;\;\;\;\;\;
 {\rm and} \nonumber \\
&&{\cal Z}_{{\bf q},\lambda} = v_{{\bf q},\lambda} 
/\sqrt{2U^2 |d\lambda_{{\bf q},\lambda}/d\omega| }
\end{eqnarray}
in the $y'$ and $z'$ directions.
Expressions for magnon energy and amplitudes in the strong-coupling limit
are given in the Appendix.

\ \\
\ \\

\section{Self-energy correction}
Due to multiple magnon emission and absorption processes 
associated with fermion motion in the AF state, 
the fermion self energy matrix in the spin-sublattice basis is obtained as 
\begin{eqnarray}
& & [\Sigma({\bf k},l,\omega)]_{\alpha\beta} \nonumber \\
&=&
U^2 \int\frac{d\Omega}{2\pi i} \sum_{\mu\nu} [\sigma_\mu] [G_{\alpha\beta}
({\bf k-q},m,\omega-\Omega)]
[\sigma_\nu] [ \chi ({\bf q},\Omega)]_{\alpha\beta}^{\mu\nu}\nonumber \\
\end{eqnarray}
in the non-crossing (rainbow) approximation. 

\begin{figure}
\vspace*{-25mm}
\hspace*{-27mm}
\psfig{figure=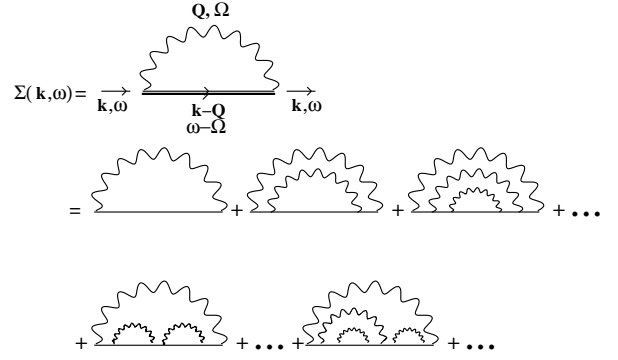,angle=-90,width=130mm}
\vspace{-30mm}
\caption{Self energy in the noncrossing approximation.
Wavy lines represent the magnon propagator.
The bare fermion-magnon interaction vertex is $U$.} 
\end{figure}

We consider the intraband contribution involving hole (electron) scattering states 
in the lower (upper) band, which redistributes the spectral function,
leaving the integrated spectral weight and the sublattice magnetization unchanged.
We obtain for the hole self energy
\begin{eqnarray}
\Sigma_{{\bf k},l} (\omega) &\equiv & \langle {\bf k},l| [\Sigma({\bf k},l,\omega)]
| {\bf k},l \rangle \nonumber \\
& = & \sum_{{\bf q},m,\lambda} \frac{|M|^2}
{\omega+\omega_{{\bf q},\lambda} - E_{{\bf k-q},m} ^- - 
\Sigma_{{\bf k-q},m} (\omega+\omega_{{\bf q},\lambda} ) }  \nonumber \\
\end{eqnarray}
where the fermion-magnon scattering matrix element
\begin{equation}
M = U \sum_{\alpha\mu} \langle {\bf k},l| \sigma_\mu | {\bf k-q},m \rangle_\alpha
\otimes | {\bf q},\lambda \rangle_{\alpha\mu} ^{\rm A} 
\end{equation}
involves the advanced magnon mode.
Substituting the sublattice structure of the fermion and magnon amplitudes,
the sum over sublattice index $\alpha$ yields
\begin{equation}
1 + e^{i(m+\lambda -l)2\pi/3} + e^{-i(m+ \lambda -l)2\pi/3} = 3\, 
\delta_{m+\lambda -l}
\end{equation}
effectively amounting to a conservation of sublattice polarization
at the fermion-magnon interaction vertex.
Therefore, the fermion-magnon scattering matrix element
reduces to a sum of the $y'$ and $z'$ fluctuation terms
\begin{eqnarray}
M = \frac{U}{\sqrt{3}}
[&-&(\alpha_{{\bf k},l}^- \beta_{{\bf k-q},m}^- - 
\beta_{{\bf k},l}^- \alpha_{{\bf k-q},m}^- )
\, {\cal Y}_{{\bf q},\lambda} \nonumber \\
&+& (\alpha_{{\bf k},l}^- \alpha_{{\bf k-q},m}^- - 
\beta_{{\bf k},l}^- \beta_{{\bf k-q},m}^- )
\, {\cal Z}_{{\bf q},\lambda}  ] \; .
\end{eqnarray}

For fermion and magnon states in the matrix element $M$, 
the sublattice-basis MBZ description of momentum 
translates to a BZ description according to the correspondence 
${\bf k},l \rightarrow {\bf k}+ l{\bf Q}$ and 
${\bf q},\lambda \rightarrow {\bf q}+ \lambda {\bf Q}$.
With this equivalence, Eq. (26) simply corresponds to momentum conservation in the BZ.
Analysis of the fermion-magnon matrix element $M$ in the strong-coupling limit
and comparison with earlier results for the $t-J$ model is discussed 
in the Appendix.
\begin{figure}
\vspace*{-50mm}
\hspace*{-70mm}
\psfig{figure=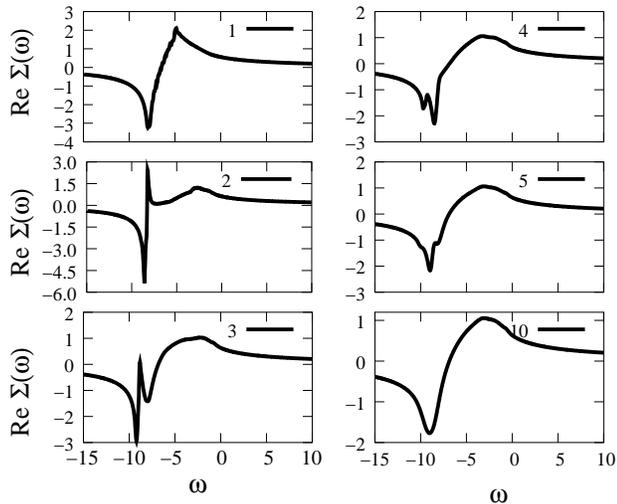,width=170mm,angle=-90}
\vspace{-55mm}
\caption{Variation of hole self energy with iterations, effectively
illustrating the role of successively higher-order magnon processes.
Here $\Delta=4$ $(U=8.8)$ and ${\bf k}=0,\; l=-1$.}
\end{figure}

It is important to note here that long-wavelength magnon modes yield 
finite contribution to the fermion-magnon scattering process
in the triangular-lattice AF, unlike the square-lattice case
where this contribution is negligible.\cite{self} 
For the square-lattice AF,
the small-$q$ contribution was suppressed because the fermion-magnon 
matrix element $M^2 \sim q$ due to destructive interference 
within sublattice summation. 
For the triangular lattice also,
for $q=0$ and $\lambda=0$ (in-plane mode), 
the fermion matrix element in Eq. (25) reduces to an expectation value
which identically vanishes for $\mu = y'$ as the spins are oriented in the $x'$ direction,
yielding similar $M^2 \sim q$ behaviour for small $q$. 
However, for the out-of-plane $z'$ fluctuation modes $(\lambda=\pm 1)$, 
the fermion matrix element is finite, resulting in $M^2 \sim 1/q$
and a finite contribution of long-wavelength modes within the 
two-dimensional $(\int q dq )$ momentum summation. 
\begin{figure}
\vspace*{-70mm}
\hspace*{-35mm}
\psfig{figure=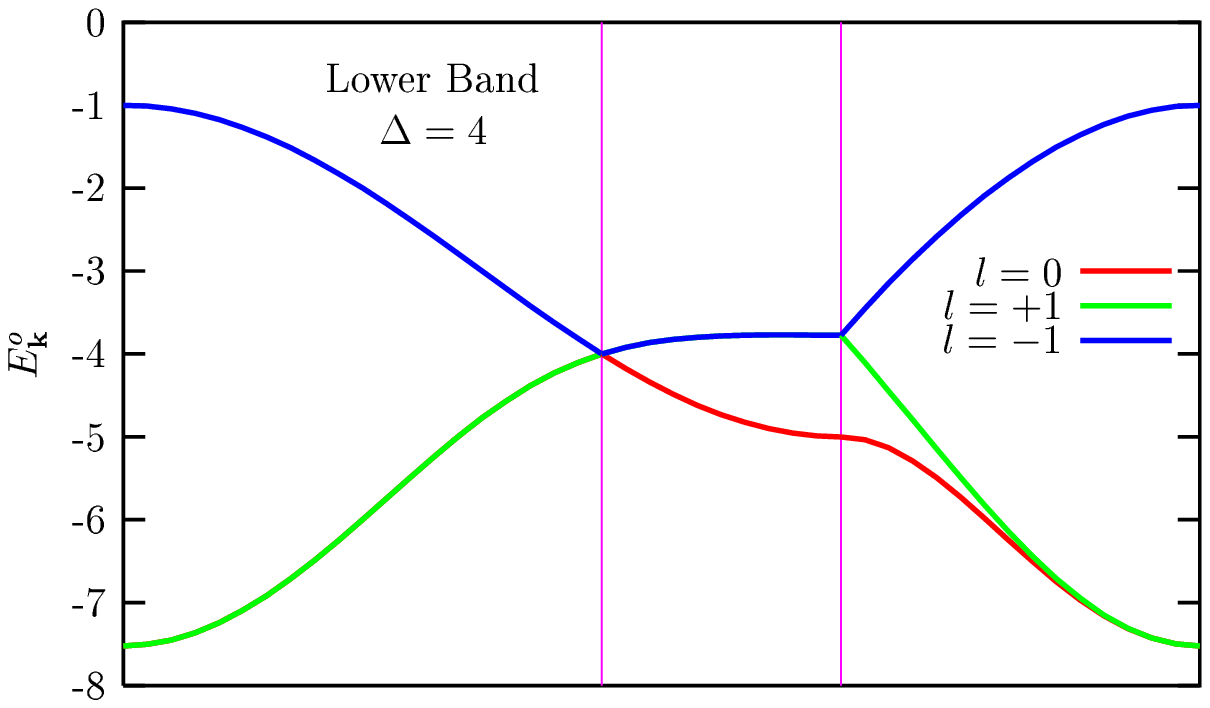,width=140mm}
\vspace{-75mm}
\vspace*{-77mm}
\hspace*{-35mm}
\psfig{figure=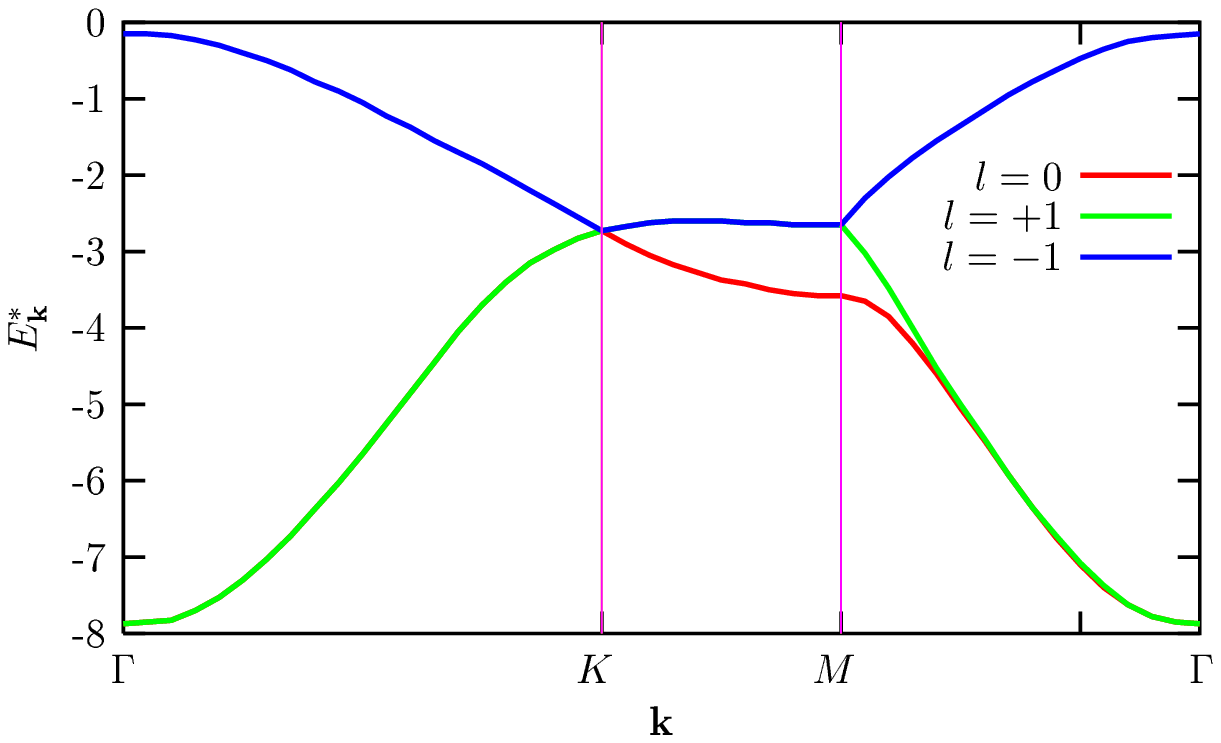,width=140mm}
\vspace{-84mm}
\caption{(color online) Quasiparticle dispersion $E_{\bf k}^\ast$ 
along different symmetry directions in the MBZ (lower panel), 
along with the HF dispersion $E_{\bf k}^0$ for comparison (upper panel).}
\end{figure}

\section{Results and Discussion} 
The self-consistent numerical evaluation of the self energy (24)
was carried out on a $30\times 40$ grid in the MBZ ${\bf k}$ space
and a frequency interval $\Delta \omega =0.025$
for $\omega$ in the range $-15 < \omega < 10$.
The self energy was iteratively evaluated,
starting with $\Sigma_{\bf k}(\omega) = 0$.
Typically, self-consistency was achieved within ten iterations 
for the lower band (Fig. 4) and fifteen iterations for the upper band (Fig. 8).

Self-energy corrections for an added hole in the 
broad, flat band near the top of the lower band are qualitatively different
from that of an added electron in the narrow, sharp peak near the bottom of the 
upper band. The low density of hole scattering states and the dominant 
band-energy denominator suppresses the hole self energy.
However, the electron self energy is significantly enhanced due to the
sharp peak and the small band energy compared with magnon energy,
resulting in the characteristic signature of string states associated
with multi-magnon processes. 
It is convenient to visualize these qualitatively different self-energy corrections 
in terms of hole (electron) motion in an effective ferromagnetic (antiferromagnetic)
spin background projected out of the $120^\circ$ spin ordering. 
\begin{figure}
\vspace*{-70mm}
\hspace*{-35mm}
\psfig{figure=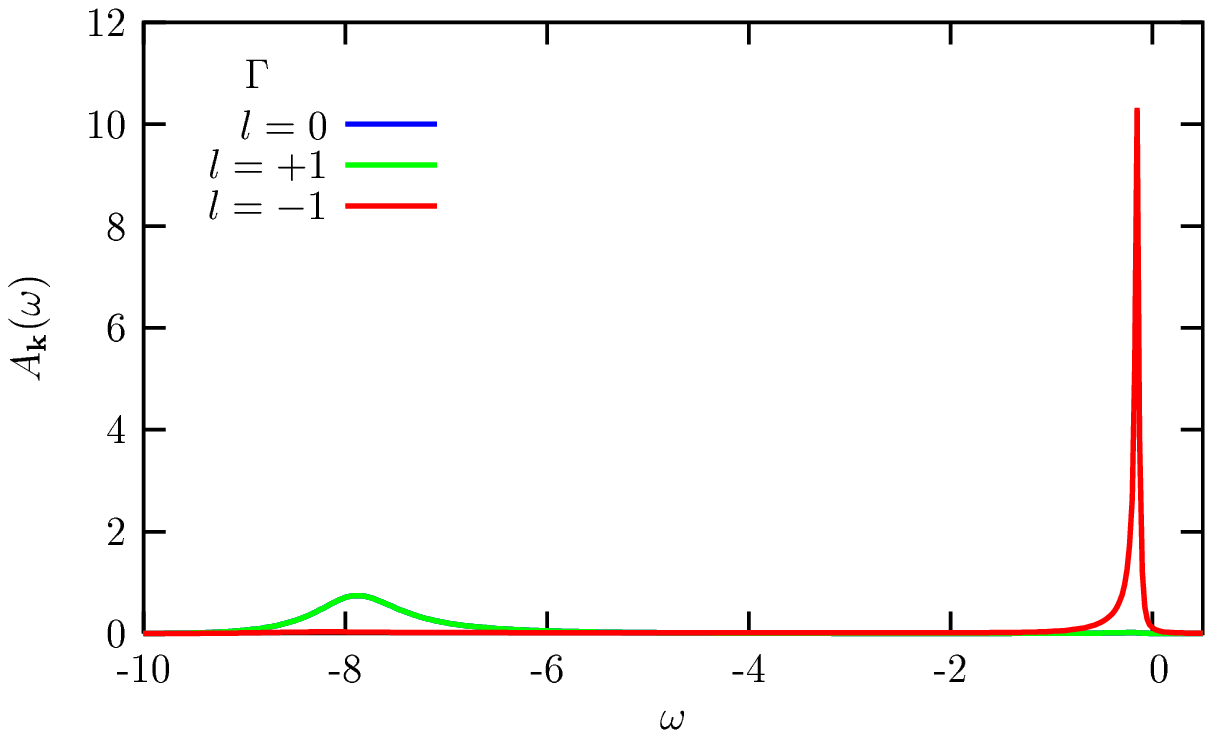,width=140mm}
\vspace{-75mm}
\vspace*{-77mm}
\hspace*{-35mm}
\psfig{figure=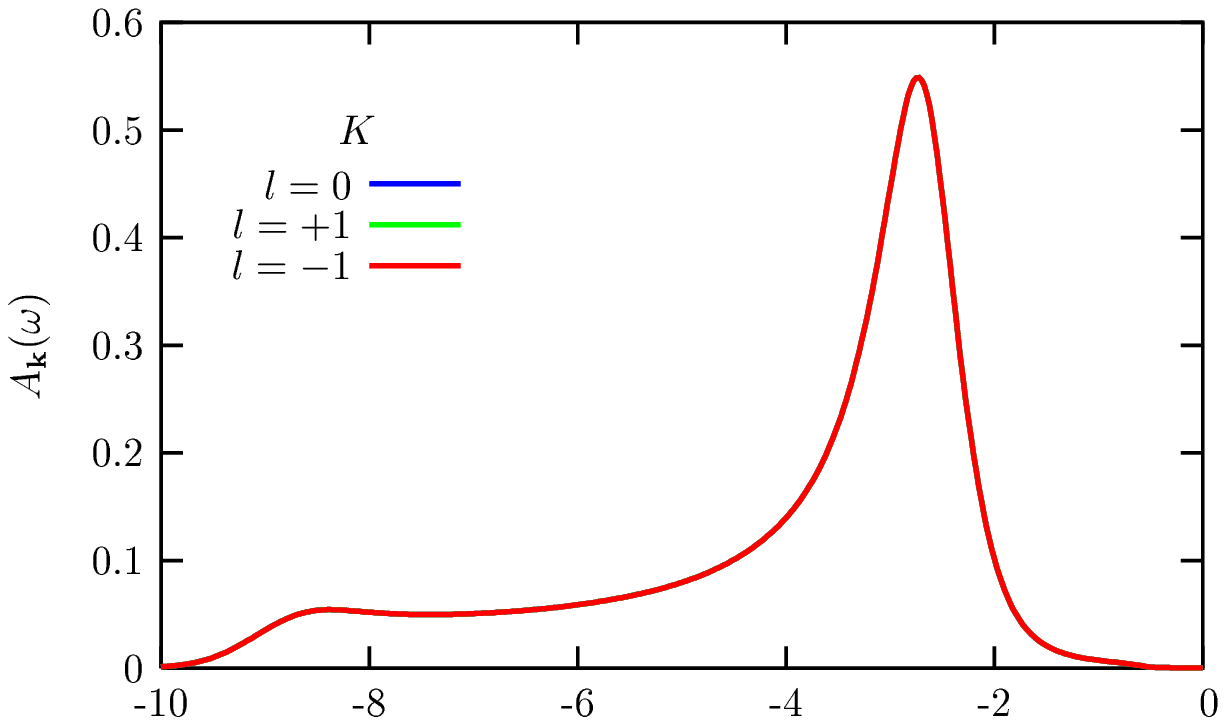,width=140mm}
\vspace{-75mm}
\vspace*{-77mm}
\hspace*{-35mm}
\psfig{figure=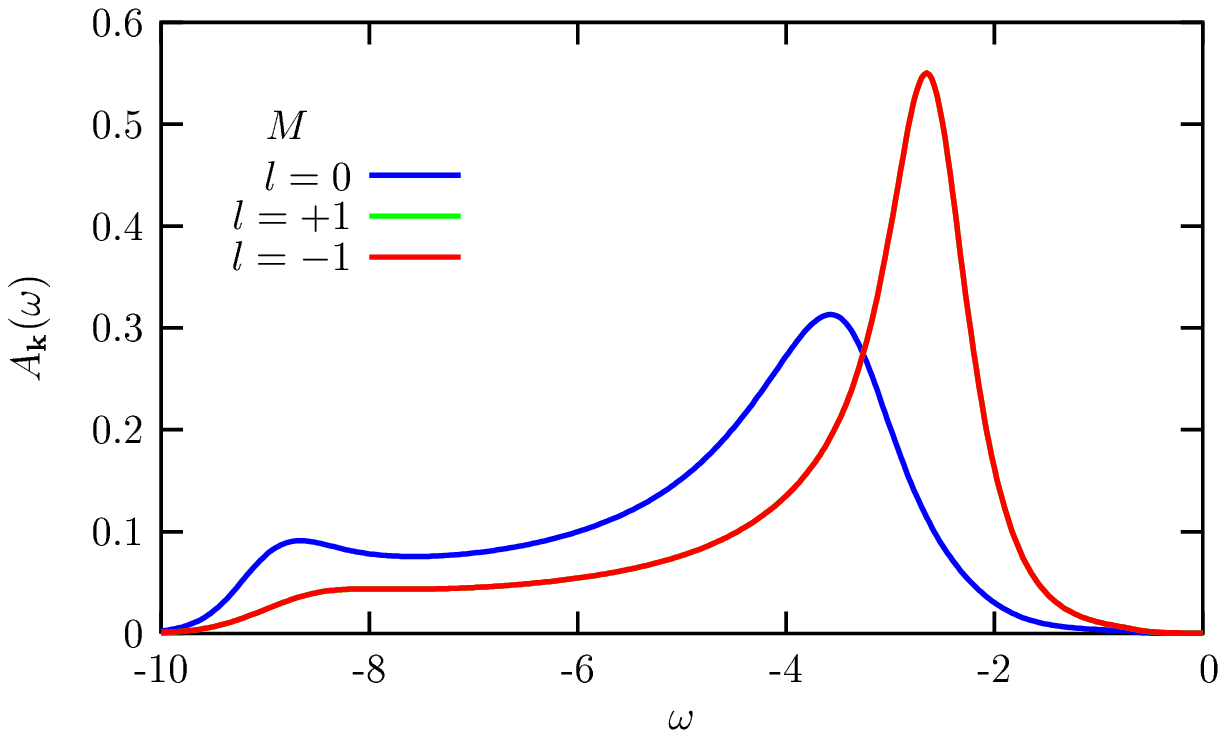,width=140mm}
\vspace{-84mm}
\caption{(color online) Hole spectral function $A_{\bf k}(\omega)$ at the special MBZ points 
$\Gamma$, K, and M. Here $\Delta=4$.}
\end{figure}

\subsection{Hole Dynamics (Lower Band)}
Figure 5 shows renormalized quasiparticle dispersion along different symmetry directions 
in the MBZ. Comparison with the HF result shows 
nearly momentum independent shift in the quasiparticle energies,
leaving the effective hole mass essentially unchanged.
States in upper portion of the band are pushed up,
while those in the lower portion are pulled down,
in accordance with the formally second-order structure of the self-energy correction. 
The quasiparticle (hole) energy is maximum (minimum) at the $\Gamma$ point 
${\bf k}=(0,0)$ for $l=-1$, corresponding to BZ momentum $-{\bf Q}$.
\begin{figure}
\vspace*{-83mm}
\hspace*{-48mm}
\psfig{figure=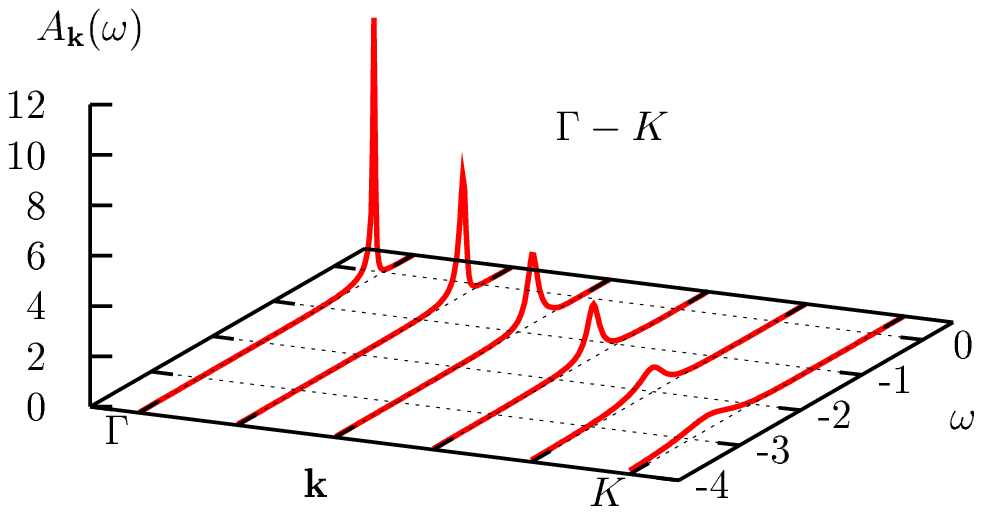,width=150mm}
\vspace{-88mm}
\vspace*{-92mm}
\hspace*{-48mm}
\psfig{figure=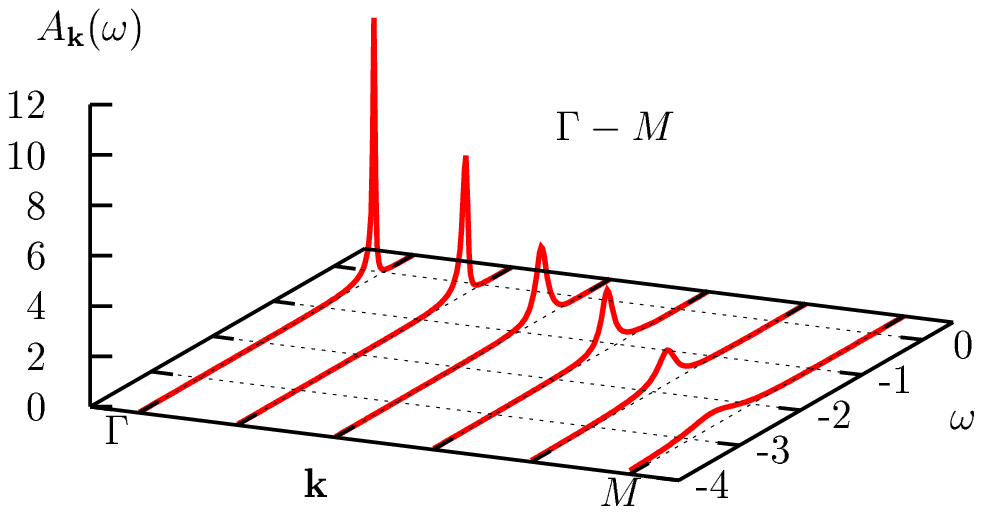,width=150mm}
\vspace{-95mm}
\caption{(color online) Hole spectral function $A_{\bf k}(\omega)$ along the $\Gamma-K$ 
and $\Gamma-M$ directions for the lowest-energy branch $l=-1$.}
\end{figure}

\begin{figure}
\vspace*{-30mm}
\hspace*{-64mm}
\psfig{figure=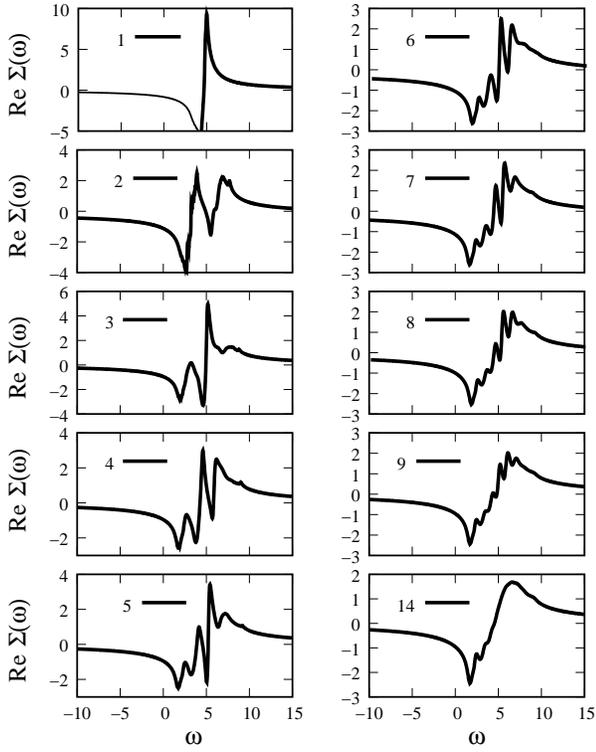,width=160mm,angle=-90}
\vspace{-33mm}
\caption{Variation of electron self energy with iterations for the upper band, 
effectively illustrating the role of successively higher-order magnon processes.
Here $\Delta=4$ and ${\bf k}=(-\frac{2\pi}{3},0)$, $l=0$.}
\end{figure}

\begin{figure}
\vspace*{-70mm}
\hspace*{-38mm}
\psfig{figure=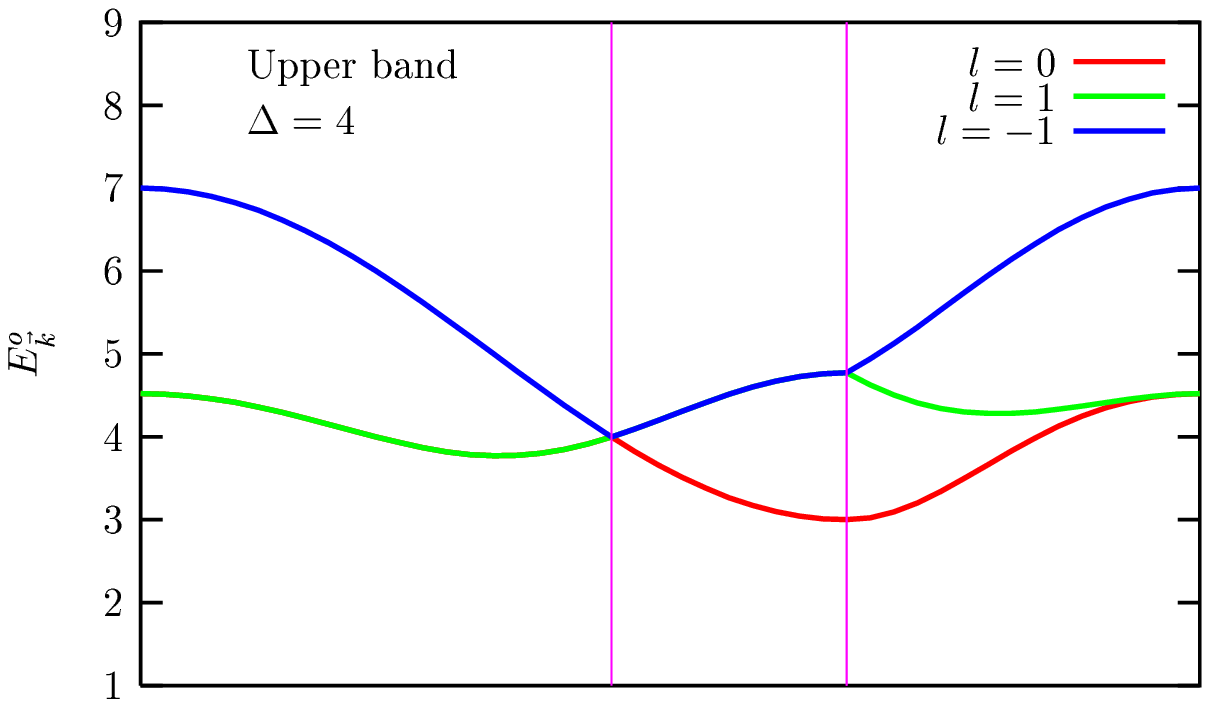,width=135mm}
\vspace{-75mm}
\vspace*{-73mm}
\hspace*{-38mm}
\psfig{figure=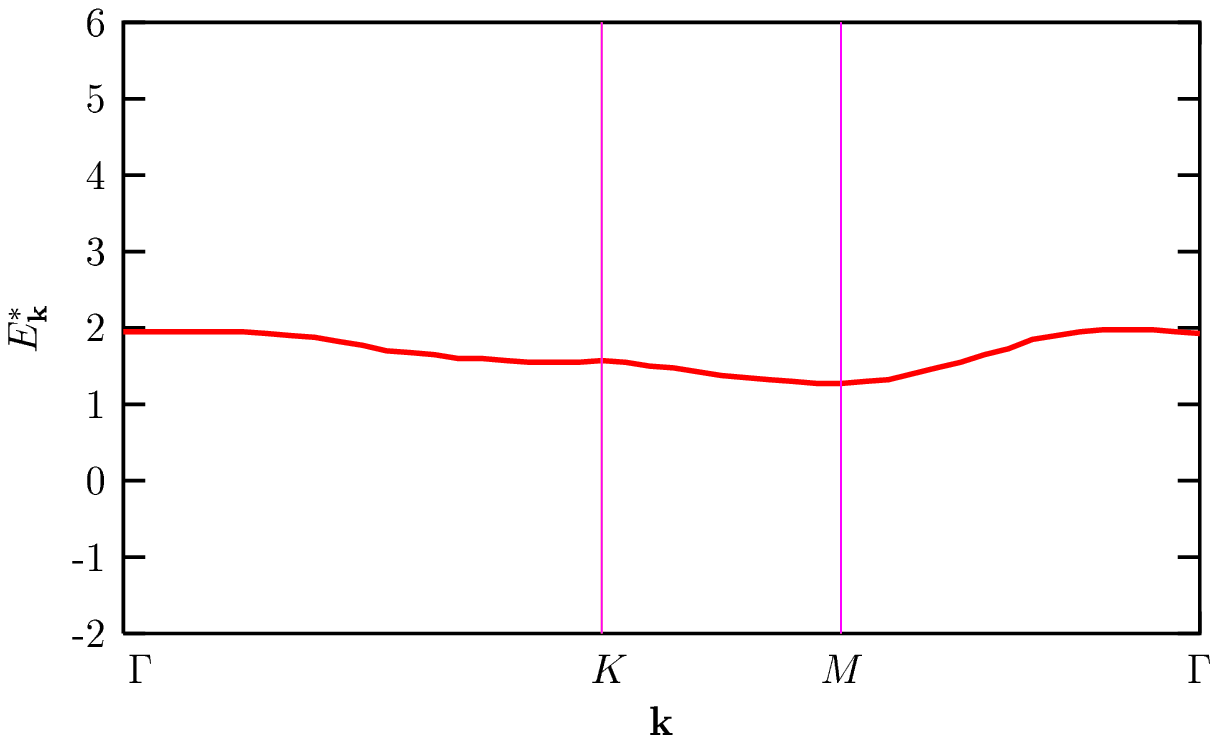,width=135mm}
\vspace{-81mm}
\caption{(color online) Quasiparticle dispersion $E_{\bf k}^\ast$ 
for the lowest-energy branch $l=0$
along different symmetry directions in the MBZ (lower panel), 
along with the HF dispersion $E_{\bf k} ^0$ for comparison (upper panel).}
\end{figure}

\begin{figure}
\vspace*{-70mm}
\hspace*{-35mm}
\psfig{figure=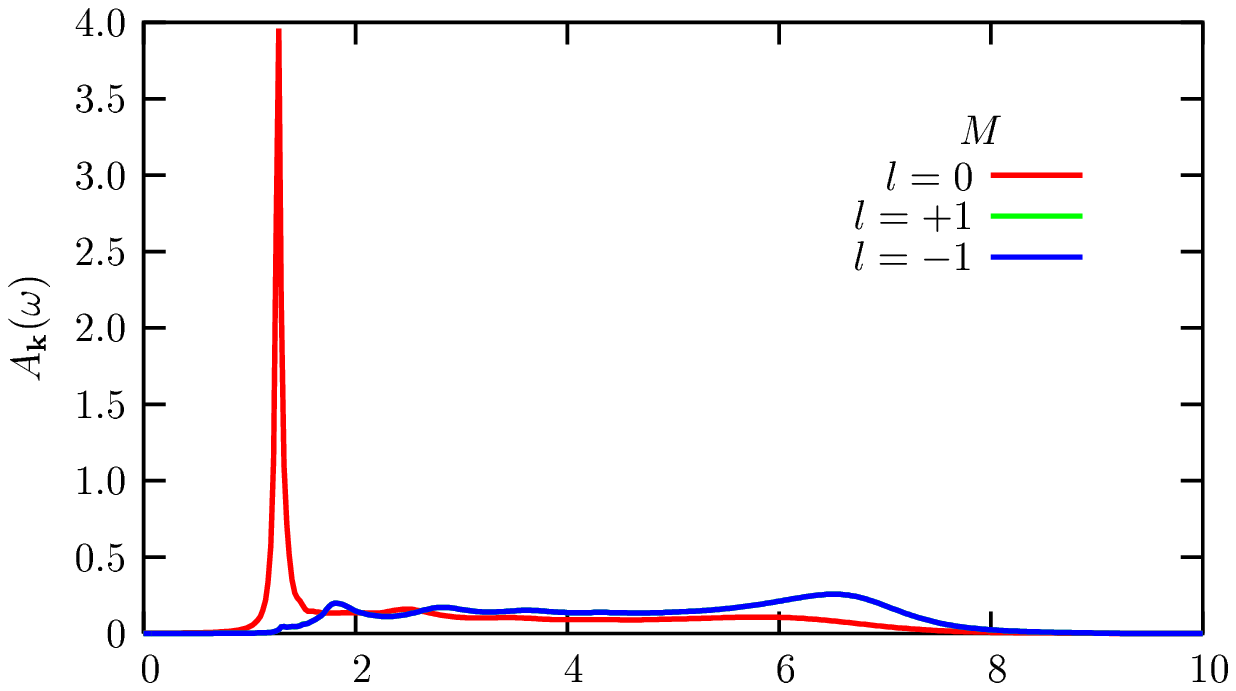,width=140mm}
\vspace{-73mm}
\vspace*{-80mm}
\hspace*{-35mm}
\psfig{figure=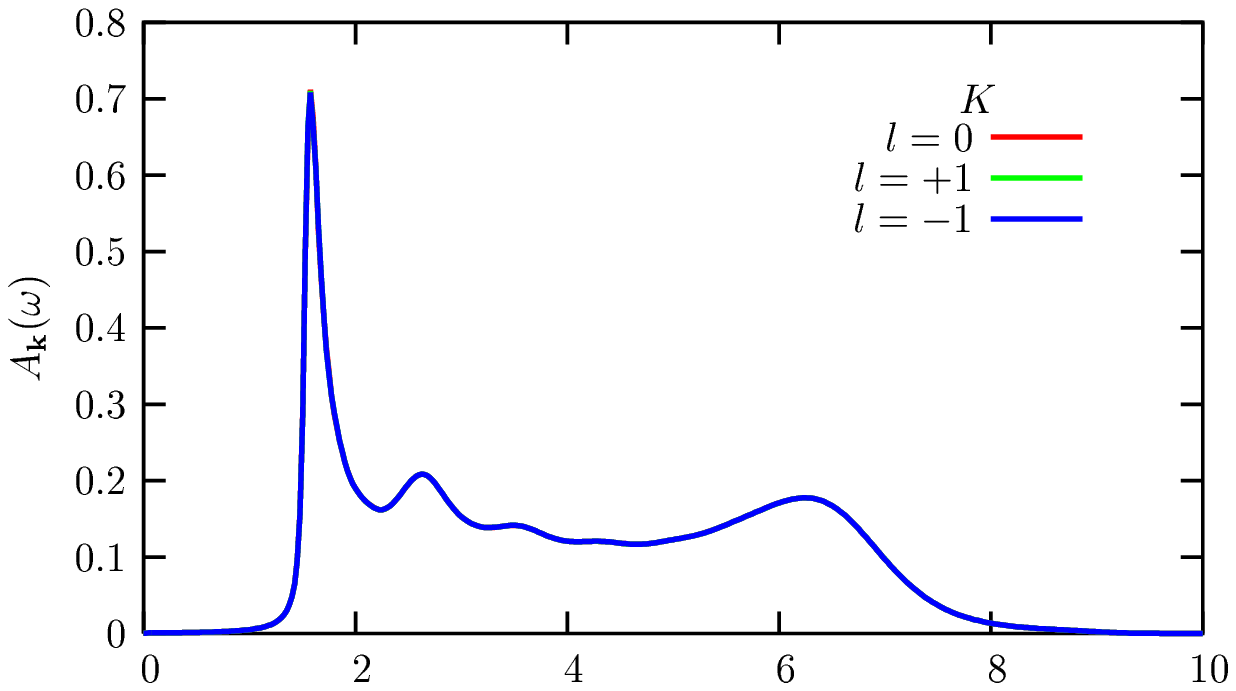,width=140mm}
\vspace{-73mm}
\vspace*{-80mm}
\hspace*{-35mm}
\psfig{figure=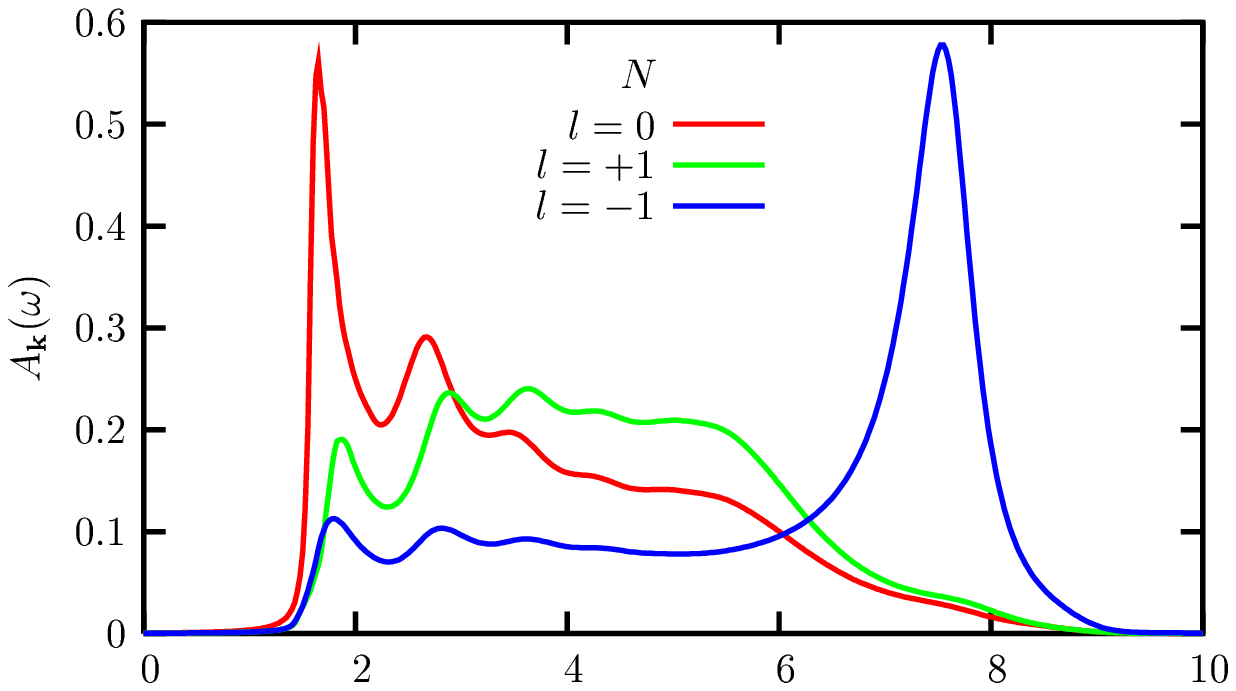,width=140mm}
\vspace{-73mm}
\vspace*{-80mm}
\hspace*{-35mm}
\psfig{figure=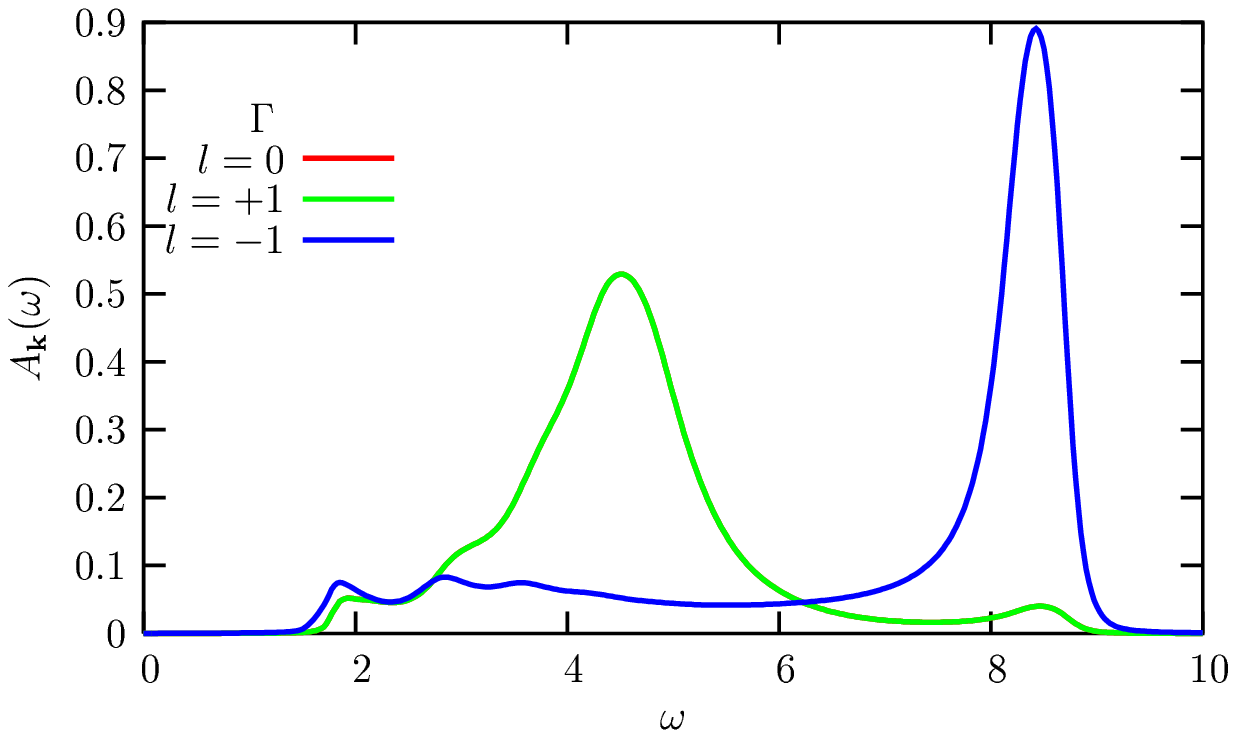,width=140mm}
\vspace{-85mm}
\caption{(color online) Electron spectral function $A_{\bf k}(\omega)$ 
at the special MBZ points for $\Delta=4$ $(U \approx 8.8)$.}
\end{figure}

Figure 6 shows the spectral function $A_{\bf k}(\omega)$ for the special points
$\Gamma$, K, and M. 
The weak self-energy correction in the broadened lower band 
of the frustrated $120^\circ$ ordered AF state
results in no visible oscillatory structure, typically associated with the 
string of broken bonds as the hole moves in the AF background.
As expected, the spectral function at the $\Gamma$ point 
${\bf k}=(0,0)$ and $l=-1$ shows a coherent quasiparticle peak, 
as this state lies at the top of the lower band. 
All three branches are degenerate at point K, as at the HF level.
The spectral functions at K and M points show well-defined peak structures,
with finite quasiparticle damping as these states lie well within the band.
The quasiparticle peak broadens and loses intensity 
along both the $\Gamma-K$ and $\Gamma-M$ directions, as seen in Fig. 7.

\subsection{Electron Dynamics (Upper Band)}
Self-energy correction in the narrow upper band is relatively large 
and self consistency is noticeably slower (Fig. 8),  
illustrating the importance of multi-magnon processes
and resulting in the characteristic oscillatory structure 
in the spectral functions (Fig. 10) associated with string states.

\begin{figure}
\vspace*{-83mm}
\hspace*{-48mm}
\psfig{figure=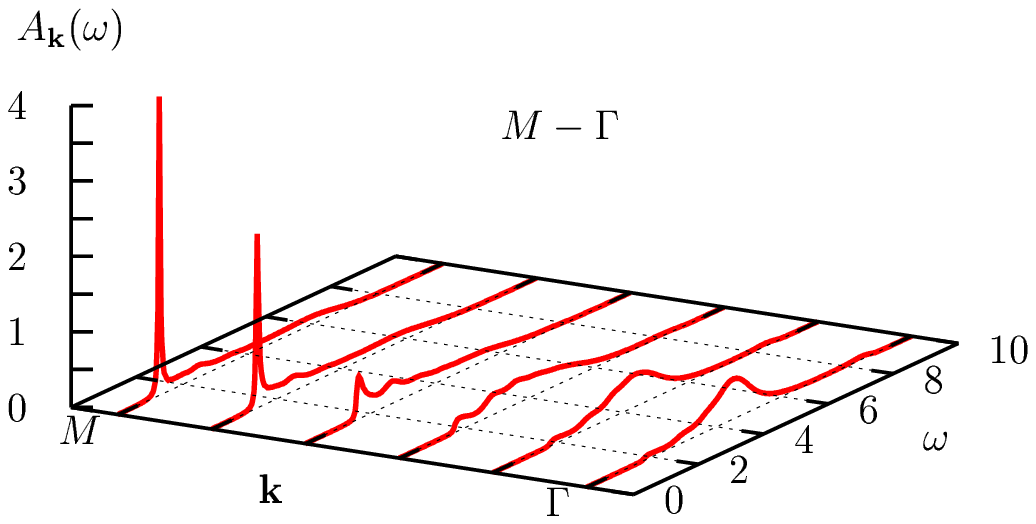,width=150mm}
\vspace{-88mm}
\vspace*{-90mm}
\hspace*{-48mm}
\psfig{figure=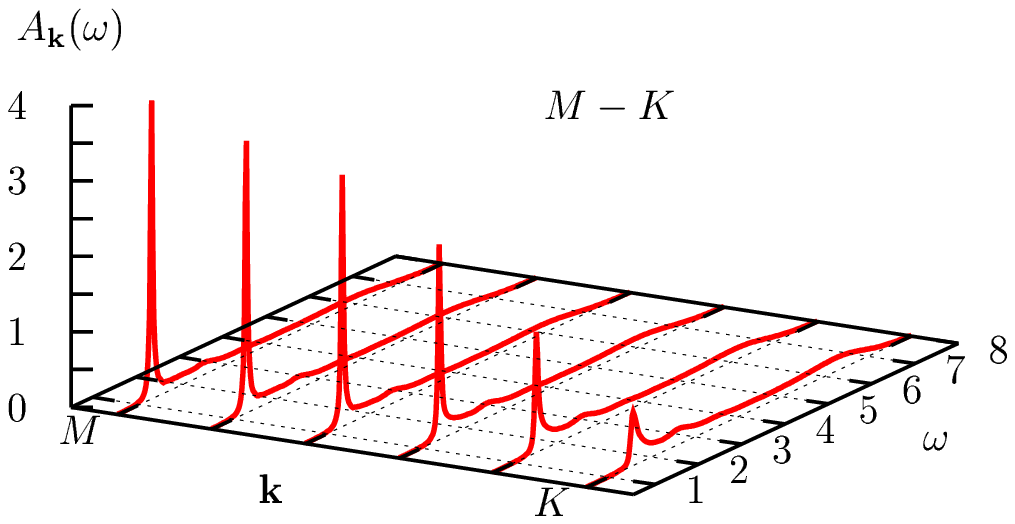,width=150mm}
\vspace{-93mm}
\vspace*{-88mm}
\hspace*{-48mm}
\psfig{figure=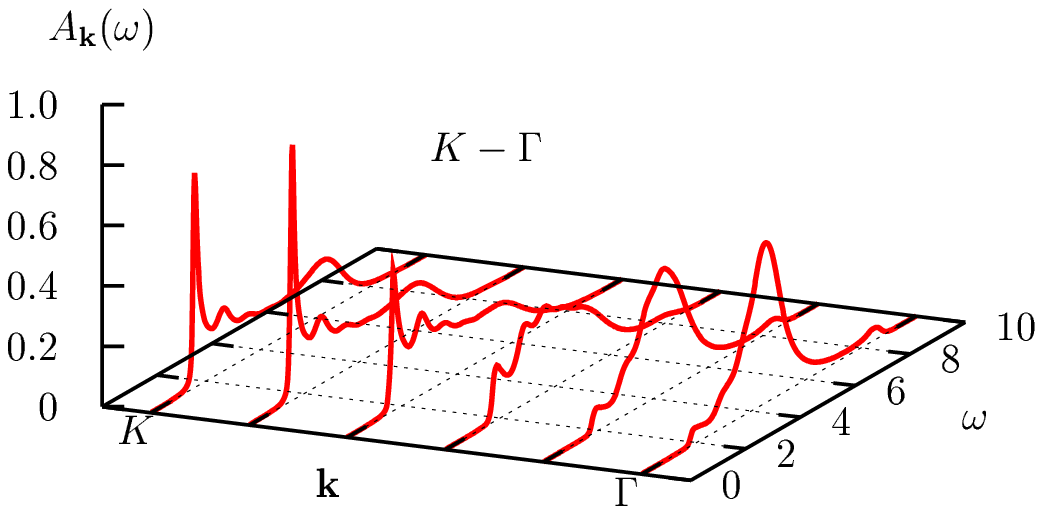,width=150mm}
\vspace{-95mm}
\caption{(color online) Evolution of the electron spectral function 
for the lowest-energy branch $l=0$ along different symmetry directions.}
\end{figure}

\begin{figure}
\vspace*{-32mm}
\hspace*{-35mm}
\psfig{figure=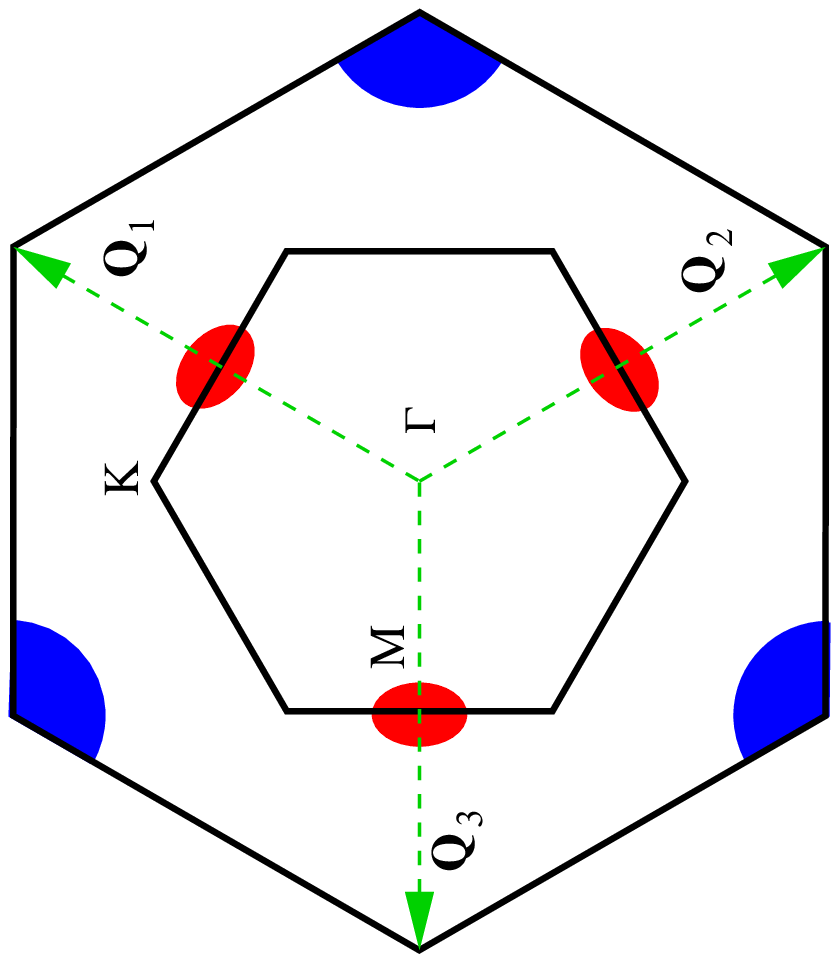,width=120mm,angle=-90}
\vspace{-35mm}
\caption{(color online) Hole (blue) and electron pockets (red) corresponding to 
lowest-energy states in the Brillouin zone. 
$\bf Q_1,Q_2,Q_3$ are equivalent AF ordering wave vectors.}
\end{figure}

Figure 9 shows the quasiparticle dispersion along the $\Gamma-K-M-\Gamma$ directions
for the lowest-energy branch $l=0$. 
The lowest-energy state at point $M$ shows a well-defined quasiparticle peak
at $\omega = 1.3$, along with a long incoherent tail, as seen in Fig. 10.
All branches are degenerate at $K$, second and third branches at $M$,
first and second branches at $\Gamma$, while at $N=(2\pi/5,0)$ 
all branches are non-degenerate, exactly as at the HF level.
\begin{figure}
\vspace*{-68mm}
\hspace*{-35mm}
\psfig{figure=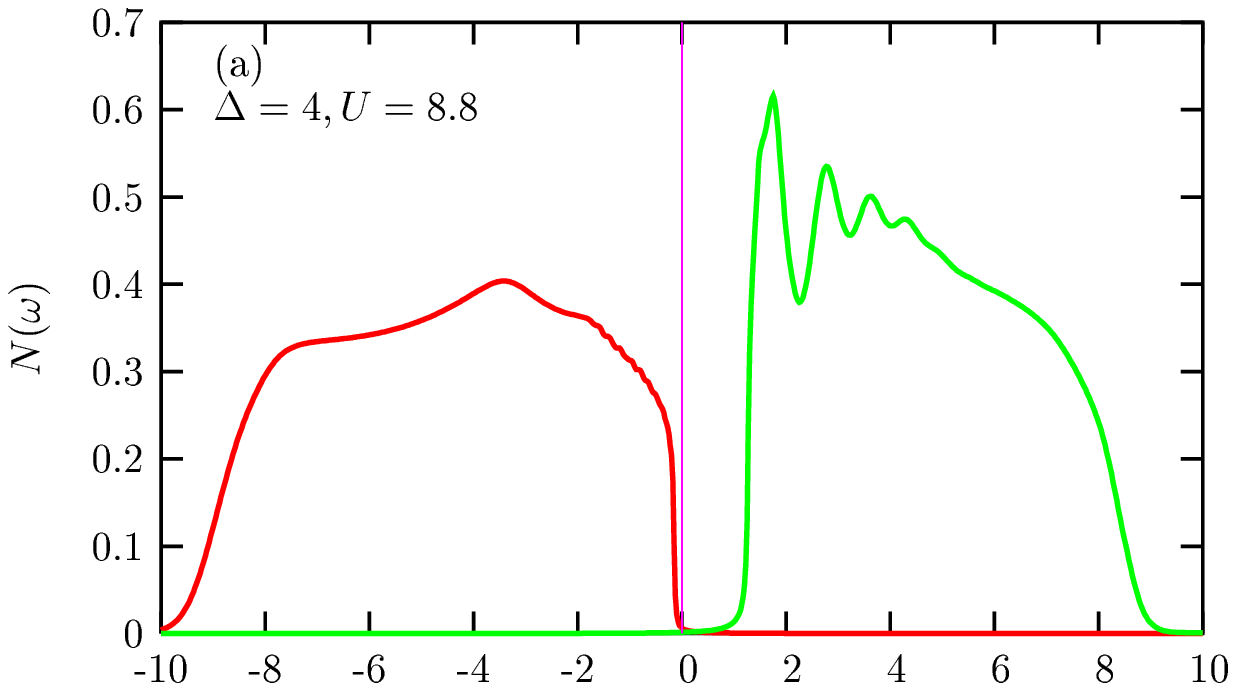,width=140mm}
\vspace{-82mm}
\vspace*{-70mm}
\hspace*{-35mm}
\psfig{figure=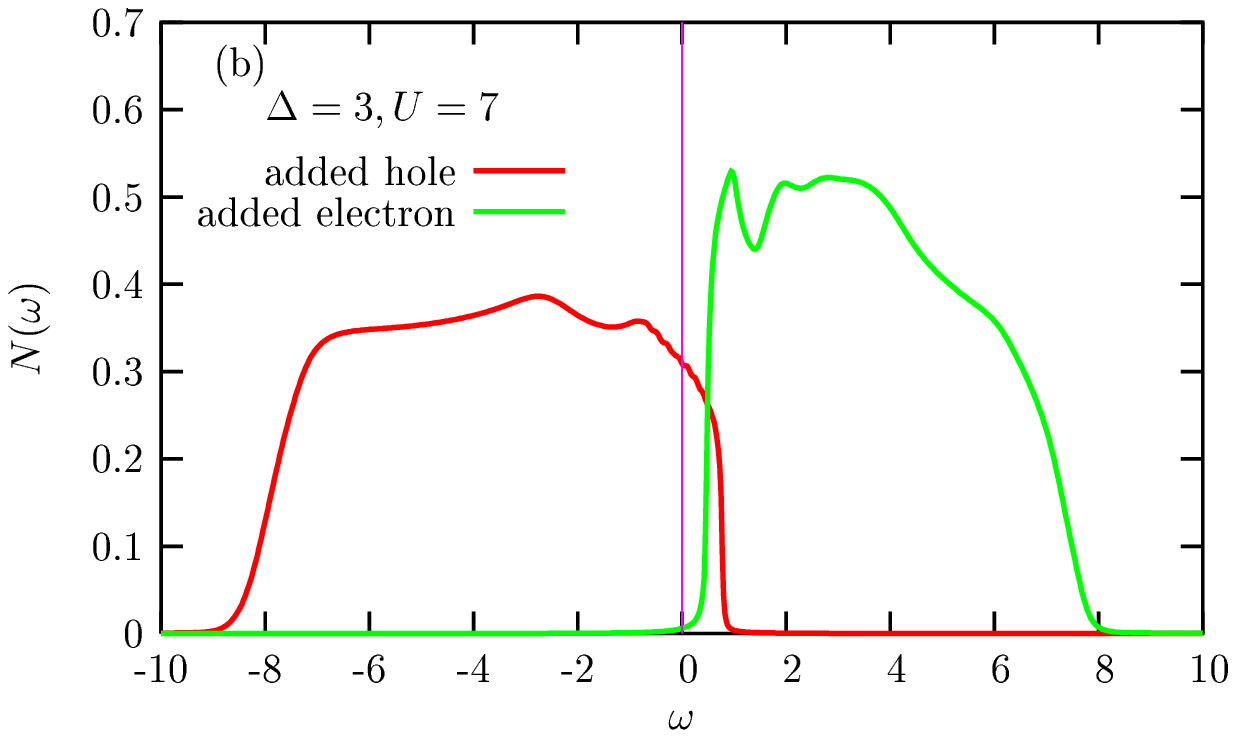,width=140mm}
\vspace{-82mm}
\caption{(color online) Renormalised density of state for one added hole and electron for 
(a) $\Delta=4$ and (b) $\Delta=3$, showing the vanishing of energy gap.}
\end{figure}

The states $K$, $\Gamma$, and $N$, which are well inside the band,
are strongly damped and yield dominantly incoherent spectral functions, 
along with small quasiparticle peaks at nearly same frequency 
$\omega \approx 2$ (Fig. 10).
It is the strong negative peak in the self energy at $\omega \approx 2$ (Fig. 8) 
which leads to nearly same quasiparticle energy for all ${\bf k}$ points,
resulting in drastically reduced quasiparticle bandwidth (Fig. 9)
and enhanced effective mass. 
Figure 11 shows that the well-defined quasiparticle peak at $M$ rapidly diminishes 
in intensity and disappears as one moves along the $M-\Gamma$ direction.
However, in the $M-K$ direction,
the quasiparticle peak is discernible in the full $k$ range.
In the $K-\Gamma$ direction,
the quasiparticle peak marginally increases and then rapidly disappears 
as one moves towards $\Gamma$. 

The lowest-energy hole and electron pockets are shown in Fig. 12.
The one-particle density of states is shown in Fig. 13 for $\Delta=4$ and $3$.
The classical-level asymmetry strongly influences the quantum corrections
and the characteristic signature of multi-magnon processes 
is dominant for the upper band corresponding to the narrow sharp 
peak at the classical level. 
As expected within the many-body expansion, 
the string-state signature of multi-magnon processes 
is more prominent for higher $U$.
The renormalized gap vanishes for $U \approx 7$, 
whereas the HF band gap $2(\Delta-2)$ vanishes at $U \approx 5$. 
The band gap is indirect, 
with the lowest-energy hole and electron states corresponding to momenta 
$(4\pi/3,0)$ and $(-2\pi/3,0)$, respectively.
 
It is interesting to note that 
in the intermediate-coupling regime of interest,
the interaction strength effectively controls the fermion-magnon scattering 
through three different aspects.
With decreasing interaction strength,
besides the explicit reduction in the fermion-magnon scattering matrix element 
due to $U$ in Eq. (25),
the classical-level fermion and magnon dispersions are also significantly modified.
Due to enhanced virtual hopping and competition with the direct hopping term
the sharp peak in the upper band DOS (Fig. 2) shifts towards higher energy.
Also, the magnon energy $\omega_M$ at momentum ${\bf Q}/2$ decreases rapidly
and vanishes at $\Delta = 2.9$.\cite{tri}
Fig. 13  shows the modifications in the upper band DOS 
due to these intermediate-$U$ effects.

\section{Conclusions}
In conclusion, we have studied the hole and electron dynamics  
in the $120^\circ$ ordered AF state
of the Hubbard model on a triangular lattice using a physically transparent 
fluctuation approach involving the dynamical spin fluctuations,
which interpolates between the weak and strong coupling limits.
Finite-$U$, double occupancy effects, neglected in earlier $t-J$ model studies, 
have been incorporated naturally in terms of classical level fermion and spin dynamics. 
Intrinsic features of the frustration-induced direct hopping dispersion 
associated with the $120^\circ$ ordering - the broad flat band and the narrow sharp peak
in the fermion DOS corresponding to lowest-energy hole and electron states -
are characteristics of ferromagnetic and antiferromagnetic ordering, respectively.
The qualitatively different self-energy corrections for hole and electron can therefore
be conveniently visualized in terms of hole (electron) motion in an 
effective ferromagnetic (antiferromagnetic) spin background 
projected out of the $120^\circ$ spin ordering. 

For an added hole in the broad lower band,
the reduced density of scattering states suppresses the fermion-magnon interaction
resulting in nearly coherent quasiparticle peak for all ${\bf k}$ states. 
No signature of string states in the spectral function 
reflects an effective F background seen by holes near the top of the lower band.
Quasiparticle dispersion shows a nearly momentum-independent shift of
hole energies, implying no mass renormalization. 
  
For an electron in the lowest-energy branch $l=0$ of the upper band,
quasiparticle peak is observed only near the MBZ boundary ($M-K$),
and rapidly vanishes away from it. 
Strong incoherent behaviour and clear signature of string states in the spectral function 
is a consequence of an effective AF background seen by electrons 
near the bottom of the upper band. 
The strong and nearly momentum-independent peak in the self energy 
leads to nearly same quasiparticle energy for all $\bf k$ points,
resulting in drastically reduced bandwidth and enhanced effective mass. 

The renormalized band gap was found to vanish for $U \approx 7$,
yielding a first-order M-I transition, as also obtained earlier 
for the frustrated square-lattice AF.\cite{self}
On the other hand, for the unfrustrated AF the band gap 
never vanishes for any finite $U$.
The vanishing of band gap at moderate $U$ for both frustrated antiferromagnets
due to frustration-induced band broadening 
thus highlights the role of frustration in M-I transition.

Finally, frustration and spin fluctuations are involved in an interesting interplay
with respect to stability of the insulating state.
Frustration generally enhances spin fluctuations and magnetic disordering. 
It will therefore be interesting to also examine the interband self-energy contribution
which reduces magnetic order due to interband spectral weight transfer,
and also widens the band gap and thereby stabilizes the insulating state.
The first-order interband contribution exactly cancels for negligible AF bandwidth,
as for the unfrustrated AF in the strong-coupling limit,\cite{self}
but will survive for finite frustration-induced bandwidth,
thus highlighting the interplay between frustration and spin fluctuations,
and providing deeper insight into the nature of the Mott insulator. 

\appendix*
\section{Fermion-magnon matrix element in the strong-coupling limit} 
We show that to leading order in the $U/t \rightarrow \infty$ limit, 
the intraband fermion matrix elements reduce to order $t/\Delta$, 
and the fermion-magnon matrix element explicitly reduces to order $t$,
as within the $t-J$ model.

For the magnon energy and magnon amplitudes in the $y'$ and $z'$ directions 
we have 
\begin{eqnarray}
\omega_{\bf q} & = & 3JS [(1-\gamma_{\bf q})(1+2\gamma_{\bf q})]^{1/2} \nonumber \\
{\cal Y}_{\bf q} ^2  &=&  
\frac{3JS}{4\omega_{\bf q}} ( 1 + 2\gamma_{\bf q} ) \nonumber \\
{\cal Z}_{\bf q} ^2  &=& 
\frac{3JS}{4\omega_{\bf q}} ( 1 - \gamma_{\bf q} ) \\
{\rm where} \;\; \gamma_{\bf q} & = & \frac{1}{3}\left 
[\cos q_x + 2 \cos \frac{q_x}{2} \cos \frac{\sqrt{3}}{2}q_y \right ] \; .\nonumber 
\end{eqnarray}

For the lower $(-)$ and upper $(+)$ bands, the fermion amplitudes 
given in Eq. (11) reduce to 
\begin{equation}
\alpha_{\bf k}^\mp \simeq  \pm \frac{1}{\sqrt{2}}
\left (1 \pm \frac{\xi_{\bf k}}{2\Delta} \right ) \;\;\; , \;\;\;
\beta_{\bf k}^\mp  \simeq  + \frac{1}{\sqrt{2}} 
\left (1 \mp \frac{\xi_{\bf k}}{2\Delta} \right ) \; .
\end{equation}
Hence in terms of the spin raising and lowering operators defined below,
the intraband fermion matrix elements 
\begin{eqnarray}
\langle \sigma^+ \rangle & \equiv &
\langle {\bf k},l| \sigma_{z'} + i \sigma_{y'} | {\bf k-q},m \rangle
\simeq \frac{\xi_{{\bf k},l}}{\Delta} \nonumber \\
\langle \sigma^- \rangle & \equiv &
\langle {\bf k},l| \sigma_{z'} - i \sigma_{y'} | {\bf k-q},m \rangle
\simeq \frac{\xi_{{\bf k-q},m}}{\Delta} 
\end{eqnarray}
for lower band states are explicitly of order $t/U$. 

In terms of corresponding magnon amplitudes 
\begin{equation}
\Phi^\pm _{{\bf q},\lambda} \equiv \sqrt{3} [ |{\bf q},\lambda \rangle_{z'} 
\pm i |{\bf q},\lambda \rangle_{y'} ]
= {\cal Z}_{{\bf q},\lambda} \pm {\cal Y}_{{\bf q},\lambda} 
\end{equation}
for the advanced mode, the fermion-magnon matrix element 
\begin{eqnarray}
\sqrt{3} M & = & U
[-i \langle \sigma_{y'} \rangle {\cal Y} + \langle \sigma_{z'} \rangle {\cal Z}] 
\nonumber \\ \nonumber \\
& = &U [\langle \sigma^+ \rangle \Phi^- + \langle \sigma^- \rangle \Phi^+ ] /2 
\nonumber \\ \nonumber \\
& = & -(\xi_{{\bf k},l} - \xi_{{\bf k-q},m}) {\cal Y}_{{\bf q},\lambda}
+ (\xi_{{\bf k},l} + \xi_{{\bf k-q},m}) {\cal Z}_{{\bf q},\lambda}  
\nonumber \\ \nonumber \\
& = & \xi_{{\bf k},l} \Phi^- _{{\bf q},\lambda} + \xi_{{\bf k-q},m} 
\Phi^+ _{{\bf q},\lambda}  \;\;\;\;\;\; ({\rm where} \;\;m=l-\lambda)
\nonumber \\
\end{eqnarray}
explicitly reduces to order $t$.

The above expression for the fermion-magnon matrix element has exactly same structure
as the result $\sqrt{3} t M(k,q)= 
\sqrt{3} t [v_{\bf q} h_{\bf k} - u_{\bf q} h_{\bf k+q} ]$ 
obtained within the $t-J$ model.\cite{dombre} 
Indeed, magnon amplitudes $u_{\bf q}$ and $v_{\bf q}$ in Ref. [17] 
exactly correspond to ${\cal Z}_{{\bf q}} \pm {\cal Y}_{{\bf q}}$,
and $(\sqrt{3}t)h_{\bf k} \equiv \sqrt{3}t \sum_{\hat{\delta}} 
\sin {\bf k}.\hat{\delta} = - \xi_{\bf k-Q}$. 
However, the Goldstone-mode contribution appears to be different.
While our fermion-magnon matrix element vanishes for the Goldstone mode 
$q \rightarrow 0,\; \lambda=0$,
for which $\gamma_{\bf q}  \rightarrow 1$ 
and the magnon amplitude ${\cal Z} \rightarrow 0$
representing rigid spin rotation about $z$ axis,
the matrix element $M({\bf k,q})$ in Ref. [17] does not vanish. 
The other two Goldstone modes $q \rightarrow 0,\; \lambda=\pm 1$, 
for which $\gamma_{\bf q}  \rightarrow -1/2$ and ${\cal Y} \rightarrow 0$,
do yield non-vanishing matrix elements with $M^2 \sim 1/q$,
although resulting in a finite contribution $\int q dq /q $ from long-wavelength modes.
Thus, for the triangular-lattice AF, 
long-wavelength modes do contribute to the fermion-magnon scattering. 



\begin{thebibliography}{05}

\bibitem{review1}
K. Kanoda, Physica C {\bf 282-287}, 299 (1997);
K. Kanoda, Hyperfine Interact. {\bf 104}, 235 (1997).

\bibitem{review2}
R. H. McKenzie, Science, {\bf 278}, 820 (1997).

\bibitem{watersup}
K. Takada {\em et al.},
Nature {\bf 422}, 53 (2003).

\bibitem{weitering}
H. H. Weitering, X. Shi, P. D. Johnson, J. Chen, N. J. Dinardo, and S. Kempa,
Phys. Rev. Lett. {\bf 78}, 1331 (1997).

\bibitem{res1}
T. Inami, Y. Ajiro, and T. Goto, 
J. Phys. Soc. Jpn. {\bf 65}, 2374 (1996).

\bibitem{resonance}
L. E. Svistov, A. I. Smirnov, L. A. Prozorova, O. A. Petrenko, L. N. Demianets,
and A. Ya. Shapiro,
Phys. Rev. B {\bf 67} 094 434 (2003). 

\bibitem{holmium}
O. P. Vajk, M. Kenzelmann, J. W. Lynn, S. B. Kim, and S.-W. Cheong,
cond-mat/0502006 (2005).

\bibitem{sw}
S. Ghosh and A. Singh, cond-mat/0506475.

\bibitem{tri}
A. Singh, Phys. Rev. B {\bf 71}, 214406 (2005).

\bibitem{kanoda}
Y. Shimizu, K. Miyagawa, K. Kanoda, M. Maesato, and G. Saito,
Phys. Rev. Lett. {\bf 91}, 107001 (2003).

\bibitem{kagawa}
F. Kagawa, T. Itou, K. Miyagawa, and K. Kanoda, 
Phys. Rev. B {\bf 69}, 064 511 (2004);
F. Kagawa, T. Itou, K. Miyagawa, and K. Kanoda, 
cond-mat/0409437 (2004).

\bibitem{pirg2}
H. Morita, S. Watanabe, and M. Imada, 
J. Phys. Soc. Jpn. {\bf 71}, 2109 (2002).

\bibitem{pirg3}
M. Imada, T. Mizusaki, and S. Watanabe, cond-mat/0307022 (2003).

\bibitem{self}
P. Srivastava and Avinash Singh,  Phys. Rev. B, {\bf 70}, 115103 (2004).

\bibitem{spectrum}
A. Singh and P. Goswami, 
Phys. Rev. B {\bf 66}, 92402 (2002).

\bibitem{capriolti}
L. Capriolti, A. E. Trumper and S. Sorella, Phys. Rev. Lett., {\bf 82}, 3899
(1999)

\bibitem{dombre}
M. Azzouz and T. Dombre, Phys. Rev. B, {\bf 53}, 402 (1996).

\bibitem{trumper}
A. E. Trumper, C. J. Gazza and L. O. Manuel, Phys. Rev. B, {\bf 69}, 184407 (2004).

\bibitem{apel}
W. Apel, H.-U. Everts and U. K$\ddot{o}$rner, Eur. Phys. J. B, {\bf 5}, 317 (1998).

\bibitem{vojta}
Matthias Vojta, Phys. Rev. B {\bf 59}, 6027 (1999). 

\bibitem{dagotto}
M. Vojta, E. Dagotto, Phys. Rev. B, {\bf 59}, 713 (1999).

\bibitem{qmc_green}
M. C. Refolio, J. M. L\'{o}pez Sancho, and J. Rubio,
cond-mat/0103459 (2001).

\end{thebibliography}
\end{document}